\documentstyle[11pt]{article}

\newcommand{\sect}[1]{ \section{#1} \setcounter{equation}{0} }

\newcommand{\Nf}{N_{\! f}}
\newcommand{\half}{\mbox{\small{$\frac{1}{2}$}}}

\newcommand{\kslash}{k \! \! \! /}

\begin{document}

\title{Large $N_{\! f}$ methods for computing the perturbative structure of
deep inelastic scattering
\footnote{Talk presented
at Fourth International Workshop on Software Engineering and Artificial
Intelligence for High Energy and Nuclear Physics, Pisa, Italy, 3rd-8th April,
1995} }
\author{J.A. Gracey, \\ Department of Mathematical Sciences, \\ University of
Durham, \\ South Road, \\ Durham,\\ DH1 3LE, \\ \\ and \\ \\ DAMTP, \\
University of Liverpool, \\ P.O. Box 147, \\ Liverpool, \\ L69 3BX, \\
United Kingdom\footnote{Present address}}
\date{}
\maketitle

\vspace{5cm}
{\bf Abstract.} The large $N_{\! f}$ self consistency method is applied to the
computation of perturbative information in the operator product expansion used
in deep inelastic scattering. The $O(1/N_{\! f})$ critical exponents
corresponding to the anomalous dimensions of the twist $2$ non-singlet and
singlet operators are computed analytically as well as the non-singlet
structure functions. The results are in agreement with recent explicit
perturbative calculations.

\vspace{-20cm}
\hspace{13.5cm}
{\bf {LTH-346}}

\newpage

\sect{Introduction}
\noindent
Quantum chromodynamics describes the phenomenology of the strong interactions.
Currently there is a need to improve our knowledge of the higher order
structure of the field theory beyond earlier one and two loop analysis. Such
three loop calculations are performed using efficient computer algebra
programmes which handle the tedious algebra of such computations. As these are
computer based it is crucial that independent checks are carried out to confirm
the final results. We report here on a method which achieves this which is the
large $\Nf$ self-consistency approach developed in \cite{1} and applied to four
dimensional gauge theories including QED and QCD.\cite{2} Essentially the
method formulates perturbation theory in an alternative way. For models which
possess parameters additional to the coupling, such as the number of
fundamental fields like $\Nf$ quarks in QCD, another expansion parameter
exists. If $\Nf$ is large then $1/\Nf$ is small and satisfies the criterion for
doing perturbative calculations. Further in this approach\cite{1} one considers
the critical behaviour of the field theory in the neighbourhood of the
$d$-dimensional Gaussian fixed point, where there is a conformal symmetry. As
Green's functions obey a simple scaling form one computes the associated
critical exponent. This encodes information on the renormalization of the
original Green's function. Then the expansion of the exponent in powers of
$\epsilon$, $d$ $=$ $4$ $-$ $2\epsilon$, gives the coefficients of the
corresponding renormalization group function or physical quantity, at that
order in $1/\Nf$. We report here on the application of the technique to the
renormalization of the physical operators which arise in the operator product
expansion of deep inelastic scattering. In particular, we focus on the twist
$2$ flavour non-singlet and singlet operators which dominate in most momentum
r\'{e}gimes.$^{3-10}$ We also discuss the development of the method to compute
$O(1/\Nf)$ information on the process dependent moments of the structure
functions based on the earlier work of Broadhurst and Kataev\cite{11,12}
inspired by \cite{13}.

\sect{Formalism}
\noindent
The twist $2$ Wilson operators whose anomalous dimensions we compute in
critical exponent form are\cite{3}
\begin{eqnarray}
{\cal O}^{\mu_1 \ldots \mu_n}_{\mbox{\footnotesize{NS}},a} &=&
\half i^{n-1} {\cal S} \bar{q} \gamma^{\mu_1} D^{\mu_2} \ldots D^{\mu_n}
T^a_{IJ} q - \mbox{trace terms} \\
{\cal O}^{\mu_1 \ldots \mu_n}_\psi &=&
\half i^{n-1} {\cal S} \bar{q} \gamma^{\mu_1} D^{\mu_2} \ldots D^{\mu_n} q
- \mbox{trace terms} \\
{\cal O}^{\mu_1 \ldots \mu_n}_G &=&
i^{n-2} {\cal S} \, \mbox{Tr} \, G^{a\,\mu_1\nu} D^{\mu_2} \ldots
D^{\mu_{n-1}} G^{a \, ~ \mu_n}_{~ \nu} - \mbox{trace terms}
\end{eqnarray}
where $q^{Ii}$ is the quark field, $1$ $\leq$ $i$ $\leq$ $\Nf$, $1$ $\leq$ $I$
$\leq$ $N_{\!c}$, $n$ is the moment, $G^a_{\mu \nu}$ $=$ $\partial_\mu A^a_\nu$
$-$ $\partial_\nu A^a_\mu$ $+$ $f^{abc} A^b_\mu A^c_\nu$, $A^a_\mu$ is the
gluon field, $1$ $\leq$ $a$ $\leq$ $N_{\!c}$, the ${\cal S}$ denotes
symmetrization of the Lorentz indices and $T^a_{IJ}$ are the group generators.
We discuss the technique to deduce the anomalous dimensions of (1)-(3) by
considering the non-singlet case first.\cite{14} To carry out a perturbative
analysis one inserts the operator into some Green's function and determines the
pole structure with respect to some regularization. In the critical point
approach\cite{1} one deduces critical exponents by using, instead of the usual
propagators, the critical ones whose structure is determined solely from
scaling arguments and dimensional analysis.\cite{1,2} So in $d$-dimensions the
asymptotic scaling forms of the quark and gluon propagators in the critical
region, $k^2$ $\rightarrow$ $\infty$, are
\begin{eqnarray}
\tilde{q}(k) & \sim & \frac{\tilde{A}\kslash}{(k^2)^{\mu-\alpha}} \\
\tilde{A}_{\nu \sigma}(k) & \sim & \frac{\tilde{B}}{(k^2)^{\mu-\beta}}
\left[ \eta_{\nu\sigma} - (1-b)\frac{k_\nu k_\sigma}{k^2} \right]
\end{eqnarray}
where $d$ $=$ $2\mu$. The fixed point is defined to be $g_c$ where $\beta(g_c)$
$=$ $0$, $g_c$ $\neq$ $0$. From the one loop $\beta$-function in
$d$-dimensions\cite{16,17,18}
\begin{equation}
\beta(g) ~=~ (d-4)g + \left[ \frac{2}{3} T(R)\Nf - \frac{11}{6}C_2(G)
\right]g^2 + O(g^3)
\end{equation}
where $\mbox{Tr}(T^aT^b)$ $=$ $T(R)\delta^{ab}$, $T^aT^a$ $=$ $C_2(R)I$,
$f^{acd}f^{bcd}$ $=$ $C_2(G) \delta^{ab}$ and $g$ $=$ $(e/2\pi)^2$ is our
dimensionless coupling constant. Then from (6) at leading order, as $\Nf$
$\rightarrow$ $\infty$,
\begin{equation}
g_c ~=~ \frac{3\epsilon}{T(R)\Nf} + O \left( \frac{1}{\Nf^2}\right)
\end{equation}
The dimension of the fields in (4) and (5) are determined from the fact that
the action is dimensionless and their anomalous pieces are defined via
\begin{equation}
\alpha ~=~ \mu - 1 + \half \eta ~~~,~~~ \beta ~=~ 1 - \eta - \chi
\end{equation}
where $\eta$ is the quark anomalous dimension and $\chi$ is the anomalous
dimension of the quark gluon vertex and each have been calculated at
$O(1/N_{\! f})$ in the Landau gauge.\cite{2} The quantities $\tilde{A}$ and
$\tilde{B}$ in (4) and (5) are the amplitudes of the fields and $b$ is the
covariant gauge parameter. As (1)-(3) are physical then their anomalous
dimensions are gauge independent.\cite{3} So calculating in an arbitrary gauge
and observing the cancellation of $b$ provides a non-trivial check on any
calculation, though the work of \cite{3,4} was carried out in the Feynman
gauge.

To proceed in the critical point analysis the regularization is introduced by
shifting $\beta$ $\rightarrow$ $\beta$ $-$ $\Delta$.\cite{19} Then the residue
of the pole in $\Delta$ when (1) is inserted in a Green's function
$\langle \bar{q} {\cal O}_{\mbox{\footnotesize{NS}}} q \rangle$ contributes to
the renormalization or exponent of the bare operator,
$\gamma^{(n)}_{\cal O}(g_c)$. The full gauge independent exponent is given by
\begin{equation}
\gamma^{(n)}_{\mbox{\footnotesize{NS}}}(g_c) ~=~ \eta ~+~ \gamma^{(n)}_{\cal O}
(g_c)
\end{equation}
Then we find the anomalous dimension exponent of (1) is
\begin{eqnarray}
\gamma^{(n)}_{\mbox{\footnotesize{NS}},1}(g_c) &=& \frac{2C_2(R)(\mu-1)^2
\eta^{\mbox{o}}_1}{(2\mu-1)(\mu-2)T(R)}
\left[ \frac{(n-1)(2\mu+n-2)}{(\mu+n-1)(\mu+n-2)} \right. \nonumber \\
&&+~ \left. \frac{2\mu}{(\mu-1)}[\psi(\mu-1+n) - \psi(\mu)] \right]
\end{eqnarray}
in arbitrary dimensions, where $\psi(x)$ is the logarithmic derivative of the
$\Gamma$-function and the subscript $1$ denotes the coefficient of $1/\Nf$
in the expansion of $\gamma^{(n)}_{\mbox{\footnotesize{NS}}}(g_c)$.

There are several checks on (10). First, when $n$ $=$ $1$, the original
operator corresponds to a symmetry generator and therefore its anomalous
dimension ought to vanish\cite{20}. It is trivial to see that
$\gamma^{(1)}_{\mbox{\footnotesize{NS}},1}(g_c)$ $=$ $0$. Second, performing
the $\epsilon$-expansion of (10) and comparing with the explicit perturbative
function $\gamma^{(n)}_{\mbox{\footnotesize{NS}}}(g)$, evaluated at (7), the
coefficients agree with the leading order two loop analytic forms given
in\cite{3,4,5}. Recently information on the {\em full} $3$-loop structure has
been provided.\cite{21} That calculation made extensive use of a computer
algebra programme to the extent that
$\gamma^{(n)}_{\mbox{\footnotesize{NS}}}(g)$ is known at $3$-loops for the even
moments up to $n$ $=$ $10$. Expanding (10) to $O(\epsilon^3)$ one obtains the
analytic expression
\begin{eqnarray}
a^{\mbox{\footnotesize{NS}}}_3 &=& \frac{2}{9}S_3(n) - \frac{10}{27}S_2(n)
- \frac{2}{27}S_1(n) + \frac{17}{72} \nonumber \\
&&-~ \frac{[12n^4+2n^3-12n^2-2n+3]}{27n^3(n+1)^3}
\end{eqnarray}
for the $3$-loop coefficient at $O(1/\Nf)$, where $S_l(n)$ $=$
$\sum_{i=1}^n1/i^l$.  Substituting for the various $n$ we record that (11) is
in exact agreement with\cite{21}. Analytic expressions at $O(1/\Nf)$ can be
obtained for the higher order coefficients.\cite{14}

\sect{Singlet Anomalous Dimensions}
\noindent
We now turn to the singlet case. As (2) and (3) have the same dimensions they
mix under renormalization which complicates the computation of their anomalous
dimensions. By contrast to (1) one has a matrix of anomalous dimensions,
$\gamma^{(n)}_{ij}(g)$.\cite{3} Its eigenvalues correspond to the anomalous
dimensions of two independent combinations of the original (bare) operators
which are of physical interest. Perturbatively $\gamma^{(n)}_{ij}(g)$ is
determined in the same way as $\gamma^{(n)}_{\mbox{\footnotesize{NS}}}(g)$ but
the operators are also included in a gluon $2$-point function $\langle A
{\cal O} A \rangle$ and it, and therefore its eigenvalues
$\gamma^{(n)}_\pm(g)$, have been computed to two loops.$^{7-10}$ To construct
the corresponding anomalous dimension exponents, one analyses the leading order
large $\Nf$ graphs at criticality as discussed before but now determines the
matrix of critical exponents. The graphs for these have been given
elsewhere\cite{7} but we note that there are several two loop graphs which
contribute at leading order in $1/\Nf$. We find the eigenvalues of
$\gamma^{(n)}_{ij}(g)$ at criticality for $n$ even, at leading order, are
\begin{eqnarray}
\gamma^{(n)}_{+}(g_c) &=& - \, 2(\mu-2) \\
\gamma^{(n)}_{-}(g_c) &=& \frac{C_2(R)\eta^{\mbox{o}}_1}{(2\mu-1)
(\mu-2)T(R)\Nf} \left[ \frac{2(\mu-1)^2(n-1)(2\mu+n-2)}{(\mu+n-1)(\mu+n-2)}
\right. \nonumber \\
&&+~ \left. 4\mu(\mu-1)[\psi(\mu-1+n) - \psi(\mu)] \right. \nonumber \\
&&-~ \left. \frac{\mu(\mu-1)\Gamma(n-1)\Gamma(2\mu)}{(\mu+n-1)(\mu+n-2)
\Gamma(2\mu-1+n)} \right. \nonumber \\
&&~~~~~ \times \left. [(n(n-1)+2(\mu-1+n))^2 \right. \nonumber \\
&&~~~~~ \left. +~ 2(\mu-2)(n(n-1)(2\mu-3+2n) + 2(\mu-1+n))] \frac{}{} \right]
\end{eqnarray}
In the critical point approach (12) and (13) follow naturally without
diagonalization as $\gamma^{(n)}_{ij}(g_c)$ is triangular at leading order. Eq.
(12) represents the anomalous dimension of the operator which has (3) as its
dominant contribution in the linear combination we mentioned, whilst (13)
corresponds to (2) being dominant. The $\Nf$ dependence of each term is
different due to the particular way in which quark loops occur at leading order
in the Green's functions. Specifically the $\epsilon$-expansion of (12)
involves only a one loop term and no subsequent terms, consistent with the
perturbative analysis. (We also note that recently the renormalization of the
gluonic sector has been re-examined to understand the role infrared divergences
play.\cite{22})

There are several checks on (13). When $n$ $=$ $2$, the operator corresponds to
the energy momentum tensor and since it is conserved its anomalous dimension
vanishes.\cite{20} Clearly, $\gamma^{(2)}_{-}(g_c)$ $=$ $0$. Secondly, using
the explicit $2$-loop matrix, $\gamma^{(n)}_{ij}(g)$, and extracting both
eigenvalues as a perturbative expansion in $g$, the coefficients of (13) to
$O(\epsilon^2)$ agree exactly.$^{7-10}$ We have also checked the cancellation
of the gauge parameter. For comparison sake we record that the three loop
leading order $1/\Nf$ coefficient of $\gamma^{(n)}_{-}(g)$ is
\begin{eqnarray}
a^\psi_3 &=& \frac{2}{9}S_3(n) - \frac{10}{27}S_2(n) + \frac{17}{72}
- \frac{2(n^2+n+2)^2[S_2(n)+S^2_1(n)]}{3n^2(n+2)(n+1)^2(n-1)} \nonumber \\
&-& 2S_1(n)[n^9+6n^8-36n^7-216n^6-552n^5-810n^4-811n^3 \nonumber \\
&& - \, 690n^2-132n+72]/[27(n+2)^2(n+1)^3(n-1)n^3] \nonumber \\
&-& [100n^{10}+682n^9+2079n^8+3377n^7+3389n^6+3545n^5+3130n^4
\nonumber \\
&& + \, 118n^3-940n^2-72n+144]/[27(n+2)^3(n+1)^4n^4(n-1)]
\end{eqnarray}

\sect{Structure Function Moments}
\noindent
A second ingredient in the Wilson expansion is the determination of the process
dependent Wilson coefficients, $C^i(q^2/m^2,g)$, $i$ $=$ $\mbox{NS}$, $\psi$ or
$G$, whose momentum evolution is controlled by the anomalous dimensions. As
these have also been computed at $O(g^3)$ and to $n$ $=$ $10$ in \cite{21} it
is equally important to have an efficient method of calculation in large $\Nf$.
By contrast with (10,12,13), which relate to the divergent part of Green's
functions, the $C^i$'s are determined from the {\em finite} part of the
amplitude after renormalization. The most efficient way to achieve this is
based on \cite{11,12} where one computes a minimal set of one or two loop
graphs. In that approach the relevant diagram is calculated in strictly four
dimensions but where the gluon line has an exponent of $1$ $+$ $\delta$ instead
of $1$. Multiplying the resultant expression, which will be a function of
$\delta$, by $e^{5\delta/3}$ the coefficients of its Taylor series in $\delta$
correspond to the perturbative coefficients of $C^i$ at $O(1/\Nf)$. This
exponential factor is necessary to ensure results are in the
$\overline{\mbox{MS}}$ scheme, and the $5/3$ arises from the finite part of the
charge renormalization. (We note that we have developed this independently of a
similar approach in \cite{23} for a different problem.) We have applied this
method to the determination of the simplest process ie the non-singlet
longitudinal Wilson coefficient in the Bjorken limit. Including the crossed
amplitude we find, at the $L$th loop,
\begin{equation}
C^{\mbox{\footnotesize{NS}}}_{\mbox{\footnotesize{long}}}(1,g) ~=~
\frac{d^L~}{d\delta^L} \! \left.
\left[ \frac{8C_2(R) e^{5\delta/3}\Gamma(n+\delta)g}{(2-\delta)(1-\delta)
(n+1+\delta)\Gamma(n)\Gamma(1+\delta)x^n} \right] \! \right|_{\delta ~=~
\frac{4}{3}\Nf T(R)g}
\end{equation}
We recall that the coupling constant expansion of
$C^{\mbox{\footnotesize{NS}}}_{\mbox{\footnotesize{long}}}(1,g)$ is $O(g)$ and
not $O(1)$. Expanding in powers of $\delta$ the coefficients agree with the
$2$-loop results of \cite{7,24,25} for all $n$. We note that the three loop
coefficient at $O(1/\Nf)$ is
\begin{eqnarray}
a^{\mbox{\footnotesize{NS}}}_{\mbox{\footnotesize{long}},3} &=&
\frac{16C_2(R)T^2(R)}{9(n+1)} \left[ \frac{203}{18} - S_2(n) + S^2_1(n)
- \frac{19(2n+1)}{3n(n+1)} \right. \nonumber \\
&&+~ \left. 2S_1(n)\left( \frac{19}{6} - \frac{1}{n} - \frac{1}{n+1}\right)
+ \frac{2}{n^2} + \frac{2}{(n+1)^2} + \frac{2}{n(n+1)} \right] \nonumber \\
\end{eqnarray}
which agrees with the even moments of \cite{21} up to $n$ $=$ $10$.

\sect{Conclusions}
\noindent
We have presented the latest results of applying the large $\Nf$
self-consistency programme to the operator product expansion of deep inelastic
scattering. Analytic results have been obtained for the higher order structure
of the anomalous dimensions of the physical operators and the moments of the
structure functions at $O(1/\Nf)$. An important feature of our results is that
we have given an insight into the (complicated) analytic structure at three and
higher loops as a function of $n$, primarily by exploiting the conformal
symmetry of the $d$-dimensional fixed point. This is a first step in gaining an
insight into the subleading $O(1/\Nf^2)$ coefficients in relation to
determining the $O(g^3)$ Altarelli-Parisi splitting functions. Further, we have
indicated how to deduce useful information compactly and efficiently, for the
moments of Wilson coefficients in $1/\Nf$. Although we have focussed on the
simplest case, the longitudinal part of the non-singlet amplitude, it ought to
be possible to adapt the approach to determine information on the singlet
structure.


\begin{thebibliography}{99}
\bibitem{1} A.N. Vasil'ev, Yu.M. Pis'mak \& J.R. Honkonen, Theor. Math. Phys.
{\bf 46} (1981), 157; {\bf 47} (1981), 291.
\bibitem{2} J.A. Gracey, Mod. Phys. Lett. {\bf A7} (1992), 1945; Int. J. Mod.
Phys. {\bf A8} (1993), 2465; Phys. Lett. {\bf B317} (1993), 415; Phys. Lett.
{\bf B318} (1993), 177.
\bibitem{3} H.Georgi \& H.D. Politzer, Phys. Rev. {\bf D9} (1974), 416; D.J.
Gross \& F.J. Wilczek, Phys. Rev. {\bf D9} (1974), 980.
\bibitem{4} E.G. Floratos, D.A. Ross \& C.T. Sachrajda, Nucl. Phys. {\bf B129}
(1977), 66; {\bf B139} (1978), 545(E).
\bibitem{5} A. Gonz\'{a}lez-Arroyo, C. L\'{o}pez \& F.J. Yndur\'{a}in, Nucl.
Phys. {\bf B153} (1979), 161.
\bibitem{6} G. Curci, W. Furmanski \& R. Petronzio, Nucl. Phys. {\bf B175}
(1980), 27.
\bibitem{7} E.G. Floratos, D.A. Ross \& C.T. Sachrajda, Nucl. Phys. {\bf B152}
(1979), 493.
\bibitem{8} A. Gonz\'{a}lez-Arroyo \& C. L\'{o}pez, Nucl. Phys. {\bf B166}
(1980), 429.
\bibitem{9} C. L\'{o}pez \& F.J. Yndur\'{a}in, Nucl. Phys. {\bf B183} (1981),
157.
\bibitem{10} W. Furmanski \& R. Petronzio, Phys. Lett. {\bf 97B} (1980), 437.
\bibitem{11} D.J. Broadhurst, Z. Phys. {\bf C58} (1993), 339.
\bibitem{12} D.J. Broadhurst \& A.L. Kataev, Phys. Lett. {\bf B315} (1993),
179.
\bibitem{13} A. Palanques-Mestre \& P. Pascual, Commun. Math. Phys. {\bf 85}
(1984), 277.
\bibitem{14} J.A. Gracey, Phys. Lett. {\bf B322} (1994), 141.
\bibitem{15} A.J. Buras, Rev. Mod. Phys. {\bf 52} (1980), 199.
\bibitem{16} D.J. Gross \& F.J. Wilczek, Phys. Rev. Lett. {\bf 30} (1973),
1343; H.D. Politzer, Phys. Rev. Lett. {\bf 30} (1973), 1346.
\bibitem{17} W.E. Caswell, Phys. Rev. Lett. {\bf 33} (1974), 244; D.R.T. Jones,
Nucl. Phys. {\bf B75} (1974), 531; E.S. Egorian \& O. V. Tarasov, Teor. Mat.
Fiz. {\bf 41} (1979), 26; O.V. Tarasov, A.A. Vladimirov \& A.Yu. Zharkov,
Phys. Lett. {\bf 93B} (1980), 429.
\bibitem{18} S.A. Larin \& J.A.M. Vermaseren, Phys. Lett. {\bf B303} (1993),
334.
\bibitem{19} A.N. Vasil'ev \& M.Yu. Nalimov, Theor. Math. Phys. {\bf 55}
(1982), 163; {\bf 56} (1983), 15.
\bibitem{20} J.C. Collins, `Renormalization' (CUP, Cambridge, 1984).
\bibitem{21} S.A. Larin, T. van Ritbergen \& J A M. Vermaseren, in `New
Computing Techniques in Physics Research III' eds K.-H. Becks \& D.
Perret-Gallix, (World Scientific, Singapore, 1994); Nucl. Phys. {\bf B427}
(1994), 41.
\bibitem{22} J.C. Collins \& R.J. Scalise, Phys. Rev. {\bf D50} (1994), 4115;
B.W. Harris \& J. Smith, `Gauge dependence of the energy-momentum tensor in
pure Yang-Mills', ITP-SB-9404, February 1994.
\bibitem{23} M. Beneke, Nucl. Phys. {\bf B405} (1993), 424.
\bibitem{24} A. Devoto, D.W. Duke, J.D. Kimel \& G.A. Sowell, Phys. Rev. {\bf
D30} (1984), 541.
\bibitem{25} D.I. Kazakov \& A.V. Kotikov, Nucl. Phys. {\bf B307} (1988), 721;
{\bf B345} (1990), 299(E).
\end{thebibliography}
\end{document}